\documentclass[letters,fleqn,useAMS,usenatbib]{mnras}
\usepackage{graphicx}
\usepackage[english]{babel}
\usepackage{amsmath}
\usepackage{amssymb}
\usepackage{natbib}
\usepackage{wasysym}
\usepackage{newtxtext,newtxmath}
\usepackage[T1]{fontenc} 

\newcommand{\se}[1]{Section~\ref{sec:#1}}

\newcommand{\fig}[1]{Fig.~\ref{fig:#1}}
\newcommand{\figs}[1]{Figs.~\ref{fig:#1}}
\newcommand{\figss}[1]{\ref{fig:#1}}
\newcommand{\Fig}[1]{Figure~\ref{fig:#1}}

\newcommand{\be}{\begin{equation}}
\newcommand{\ee}{\end{equation}}
\newcommand{\bea}{\begin{eqnarray}}
\newcommand{\eea}{\end{eqnarray}}

\newcommand{\msun}{{\rm M}_\odot}
\newcommand{\Msun}{M_\odot}

\newcommand{\ifm}[1]{\relax\ifmmode#1\else$\mathsurround=0pt #1$\fi}
\newcommand{\kms}{\ifmmode\,{\rm km}\,{\rm s}^{-1}\else km$\,$s$^{-1}$\fi}

\newcommand{\kpc}{\,{\rm kpc}}
\newcommand{\pc}{\,{\rm pc}}

\newcommand{\Myr}{\,{\rm Myr}}
\newcommand{\K}{\,{\rm K}}

\newcommand{\ltsima}{$\; \buildrel < \over \sim \;$}
\newcommand{\lsim}{\lower.5ex\hbox{\ltsima}}
\newcommand{\gtsima}{$\; \buildrel > \over \sim \;$}
\newcommand{\gsim}{\lower.5ex\hbox{\gtsima}}

\def\la{\langle}

\def\M*{M_{\rm *}}

\def\Pi{\varpi_{_{\rm I}}}

\usepackage{xcolor}

\title[Compressive Turbulence and Giant Clumps]{Formation of Giant Clumps in High-$z$ Disc Galaxies by Compressive Turbulence}
\author[N. Mandelker et al.]{Nir Mandelker,$^{1}$\thanks{E-mail: nir.mandelker@mail.huji.ac.il} 
Omry Ginzburg,$^{1}$ Avishai Dekel,$^{1,2}$ 
Frederic Bournaud,$^{3}$ 
\newauthor
Mark R. Krumholz,$^{4,5}$ Daniel Ceverino$^{6,7}$ and Joel Primack$^{8}$ \\
$^{1}$Centre for Astrophysics and Planetary Science, Racah Institute of Physics, The Hebrew University, Jerusalem 91904, Israel \\
$^{2}$SCIPP, University of California, Santa Cruz, CA 95064, USA \\
$^{3}$AIM, CEA, CNRS, Universit´ e Paris-Saclay, Universit´ e Paris Diderot, Sorbonne Paris Cit´ e, 91191 Gif-sur-Yvette, France \\
$^{4}$Research School of Astronomy and Astrophysics, Australian National University, Canberra, ACT 2611, Australia\\
$^{5}$Australian Research Council Centre of Excellence for All Sky Astrophysics in 3 Dimensions (ASTRO 3D), Australia\\
$^{6}$Departamento de Fisica Teorica, Modulo 8, Facultad de Ciencias, Universidad Autonoma de Madrid, 28049 Madrid, Spain\\
$^{7}$CIAFF, Facultad de Ciencias, Universidad Autonoma de Madrid, 28049 Madrid, Spain\\
$^{8}$Department of Physics, University of California, Santa Cruz, CA, 95064, USA\\
}

\begin{document}

\date{Accepted --. Received --; in original form --}

\pagerange{\pageref{firstpage}--\pageref{lastpage}} \pubyear{2018}

\maketitle

\label{firstpage}

\begin{abstract}
\noindent We address the formation of giant clumps in violently unstable gas-rich disc galaxies at cosmic noon. While these are commonly thought to originate from gravitational Toomre instability, cosmological simulations have indicated that clumps form even in regions where the Toomre $Q$ parameter is well above unity, which should be stable according to linear Toomre theory \citep{Inoue.etal.16}. Examining one of these cosmological simulations, we find that it exhibits an excess in compressive modes of turbulence with converging motions. The energy in converging motions within proto-clump regions is $\sim 70\%$ of the total turbulent energy, compared to $\sim 17\%$ expected in equipartition. When averaged over the whole disc, $\sim 32\%$ of the turbulent energy is in converging motions, with a further $\sim 8\%$ in diverging motions. Thus, a total of $\sim 40\%$ of the turbulent energy is in compressive modes, with the rest in solenoidal modes, compared to the $(1/3)-(2/3)$ division expected in equipartition. By contrast, we find that in an isolated-disc simulation with similar properties, resembling high-$z$ star-forming galaxies, the energy in the different turbulence modes are in equipartition, both in proto-clump regions and over the whole disc. We conclude that the origin of the excessive converging motions in proto-clump regions is external to the disc, and propose several mechanisms that can induce them. This is an additional mechanism for clump formation, complementary to and possibly preceding gravitational instability. 

\end{abstract}

\begin{keywords}
instabilities -- methods: numerical -- hydrodynamics -- galaxies: evolution -- galaxies: formation
\end{keywords}

\section{Introduction} \label{sec:intro}

Roughly $60\%$ of $\M*>10^{9}\Msun$ star-forming galaxies (SFGs) at $z\sim (1.5-3)$ are clumpy, with $\kpc$-scale clumps accounting for a few percent of their mass and $\gsim 10\%$ of their SFR \citep[e.g.][]{Guo.Y.etal.15,Shibuya.etal.16,Sattari.etal.23}. While a small fraction of these clumps may be merging galaxies \citep{Mandelker.etal.14,Zanella.etal.19,Husko.etal.23}, the majority of them are understood to have formed in situ through violent disc instabilities (VDI) \citep[e.g.][]{Noguchi.99,Elmegreen.etal.07,Bournaud.etal.07,Genzel.etal.08,Bournaud.Elmegreen.09,Dekel.etal.09b,Ceverino.etal.12,Mandelker.etal.14,Mandelker.etal.17}. VDI and clumpy star-formation are a robust feature of SFGs at $z\sim 2$, in both observations and simulations. They are thought to play a key role in mass flow to the galactic centre and the growth of bulges \citep{Elmegreen.etal.08,Dekel.etal.09b,Ceverino.etal.15,Zolotov.etal.15}, the formation of dark matter cores \citep{Ogiya.Nagai.22}, the chemical compositions of the thin and thick discs and the bulge \citep{Clarke.etal.19,Debattista.etal.23}, globular cluster formation \citep{Shapiro.etal.08,Mandelker.etal.17}, and turbulence driving in galactic discs \citep{Bournaud.etal.09,Cacciato.etal.12b,Krumholz.etal.18,Ubler.etal.19,Ginzburg.etal.22}. The question of whether or not clumps survive feedback and migrate towards the galactic center as bound structures can constrain models of stellar feedback \citep{Mandelker.etal.17,Oklopcic.etal.17,Dekel.etal.22,Dekel.etal.23}. Low-$z$ analogues of high-$z$ clumpy discs have been observed in very gas rich systems \citep{Fisher.etal.17a,Fisher.etal.17b,Mehta.etal.21}, while analogues at cosmic dawn may be related to recent observations of extremely bright galaxies with JWST \citep{Dekel.etal.23b}.

\smallskip
The in situ formation of giant clumps is commonly attributed to Toomre instability \citep{Toomre.64,Dekel.etal.09b}, where small fluctuations collapse to clumps as their self-gravity ($\Sigma$) overcomes turbulent pressure ($\sigma$) and rotational support ($\kappa$), namely once $Q\propto \sigma\kappa/\Sigma<1$. Fragmentation into massive ($M_{\rm c}\gsim 10^8\msun$) long lived clumps is commonly seen in simulations of isolated galaxy discs, especially when the gas fraction is greater than $\sim 50\%$ \citep{Fensch.bournaud.21}, though Toomre-like instabilities manifest on multiple scales even when the gas fraction is as low as $\sim 20\%$ \citep{Renaud.etal.21}. Other studies of idealized isolated galaxies suggest that while linear Toomre instability (or a related spiral arm instability, \citealp{Inoue.yoshida.18,Inoue.yoshida.19}) is responsible for the initial fragmentation, this occurs on scales smaller than the observed giant clumps which are themselves agglomerations of many small sub-clumps, either physically bound or in projection \citep[e.g][]{Romeo.Agertz.14,Agertz.etal.15,Behrendt.etal.15,Tamburello.etal.15,Benincasa.etal.19}. However, the Toomre $Q$ parameter is a \textit{linear} concept applicable to \textit{axisymmetric} systems, while the violently unstable discs at high redshift are far from being either linear or axisymmetric. Using the {\tt VELA} suite of cosmological simulations \citep{Ceverino.etal.14}, which produce clumps with properties consistent with observations \citep{Mandelker.etal.17,Guo.Y.etal.18}, \citet{Inoue.etal.16} performed a detailed two-component (gas and stars) analysis of the Toomre $Q$ parameter, following \citet{Romeo.Wiegert.11}. They found, surprisingly, that massive clumps sometimes form in regions where $Q>3$, which are expected to be stable to linear perturbations.


\smallskip
Several potential explanations for this have been proposed, such as the importance of treating different gas phases separately in the $Q$ analysis \citep{Renaud.etal.21}, the multi-scale nature of $Q$ and of disc instabilities more generally leading to small-scale fragmentation which then aggregate into larger clumps \citep{Romeo.Agertz.14,Agertz.etal.15,Benincasa.etal.19}, rapid decay of turbulence due to cooling and dissipation \citep{Elmegreen.11}, or various non-axisymmetric instabilities \citep{Toomre.81,Inoue.etal.16,Inoue.yoshida.18,Inoue.yoshida.19}. However, as discussed in \citet{Inoue.etal.16}, none of these seem sufficient to explain the simulation results. Rather, they proposed that despite the results from isolated galaxy simulations, VDI in realistic disc galaxies is not a simple linear instability, but rather a highly perturbed configuration stimulated by a non-linear process. 

\smallskip
In this letter, we present a preliminary exploration of this possibility focusing on the nature of turbulence in simulated discs. In particular, we explore whether clump formation is associated with an enhancement in the ratio of compressive to solenoidal (vortical) modes of turbulence. Previous work has shown that such an enhancement can be associated with starbursts and the formation of stellar clusters \citep{Renaud.etal.15,Renaud.etal.15b,Renaud.etal.22,Grisdale.etal.17}. We explore this question using both cosmological and isolated galaxy simulations, to examine whether the same mechanisms of clump formation are at play in both cases. 
We introduce the simulations and our methods of clump identification and turbulence decomposition in \se{sim}. In \se{results} we present our results, and we discuss them and conclude in \se{concl}.


\section{Method}
\label{sec:sim}

\subsection{Cosmological Simulation}
\label{sec:cosmo_sim}
We analyze one galaxy from the \texttt{VELA} suite of cosmological zoom-in simulations \citep{Ceverino.etal.14}, dubbed V07 (see table 1 of \citealp{Mandelker.etal.17}). The simulation utilised the \texttt{ART} code \citep{kravtsovART97,kravtsov03,Ceverino.Klypin.09}, which follows the evolution of a gravitating N-body system and the Eulerian gas dynamics with an adaptive mesh refinement (AMR) maximal resolution of $17.5-35$ physical $\pc$ at all times. The dark-matter particle mass is $8.3\times 10^4\msun$ and the minimal mass of stellar particles is $10^3\msun$. The code incorporates gas and metal cooling, UV-background photoionization and self-shielding in dense gas, stochastic star-formation, thermal feedback and metal enrichment from supernovae and stellar winds \citep{Ceverino.etal.10,Ceverino.etal.12} and feedback from radiation pressure. Further details regarding the 
simulation method can be found in \citet{Ceverino.etal.14} and \citet{Mandelker.etal.17} (models \texttt{RadPRE} and \texttt{RP}, respectively).

\smallskip
The main VDI phase in V07 lasts from $z\sim 2.5-1$, following a dramatic phase of ``wet-compaction'' to a star-forming ``blue-nugget'' \citep{Zolotov.etal.15}. During this time, the disc radius\footnote{See \citet{Mandelker.etal.14,Mandelker.etal.17} for details on how $R_{\rm d}$ and other disc properties are defined in the simulation.} steadily increases from $R_{\rm d}\sim (9-20)\kpc$, its thickness from $H_{\rm d}/R_{\rm d}\sim (0.1-0.2)$, and the disc crossing time from $t_{\rm d}\sim R_{\rm d}/V_{\rm circ}\sim (30-60)\Myr$. The baryonic disc mass increases from $M_{\rm d}\sim (2.7-6.5)\times 10^{10}\msun$, while the disc gas fraction decreases from $f_{\rm g}\sim(0.28-0.13)$. The ``cold'' mass fraction in the disc, referring to cold gas ($T<1.5\times 10^4\K$) and young stars ($<100\Myr$), declines from $f_{\rm cold}\sim (0.4-0.14)$. V07 was analyzed in \citet{Inoue.etal.16} where it was shown that many giant clumps formed in regions with $Q>3$ and even as high as $Q\sim 10$. The properties of the clumps were analyzed in detail in \citet{Mandelker.etal.17} and their temporal evolution in \citet{Dekel.etal.22}. Both of these suggest that the simulated clumps are consistent with numerous observational constraints \citep[see also][]{Guo.Y.etal.18}. As in \citet{Inoue.etal.16} and \citet{Dekel.etal.22}, we analyze the simulation with roughly 5 outputs per disc crossing time. 

\smallskip
The left-hand panel of \fig{face_on} shows a face-on projection of the gas surface density at $z\sim 2.5$, near the onset of the VDI phase. The disc is fed by two dense streams from the top and the bottom, which join the disc near its radius at $R_{\rm d}\sim 12\kpc$, marked by a white circle. Clump formation tends to occur in a thick ring near this radius \citep{Dekel.etal.20}. 

\subsection{Isolated Simulation}
\label{sec:iso_sim}

In addition to the cosmological simulation, we also analyse and idealised isolated disc galaxy simulation that serves as a control case. This simulation is similar to those described in \citet[][e.g. Model G2]{Bournaud.etal.14}, and in \citet{Perret.etal.14}. The initial galaxy mass is $3.2 \times 10^{10}\msun$ with an initial gas fraction of $65\%$, a disk half-mass radius of $3.5\kpc$, and a disc-to-total ratio of $\sim 0.6$ inside the half-stellar-mass radius. The simulation uses the \texttt{RAMSES} code \citep{Teyssier.02}, and employs an adaptive-mesh grid with a maximal AMR resolution of $7.25\pc$. The stellar feedback scheme includes supernovae feedback with thermal energy injection and a sub-grid model for non-thermal processes \citep[following][]{Teyssier.etal.13}, as well as radiation pressure and photo-ionization from young massive stars \citep[following][]{Renaud.etal.13}. 


\begin{figure*}
\begin{center}
\includegraphics[trim={0.2cm 0.0cm 0.2cm 0.2cm}, clip, width =0.97 \textwidth]{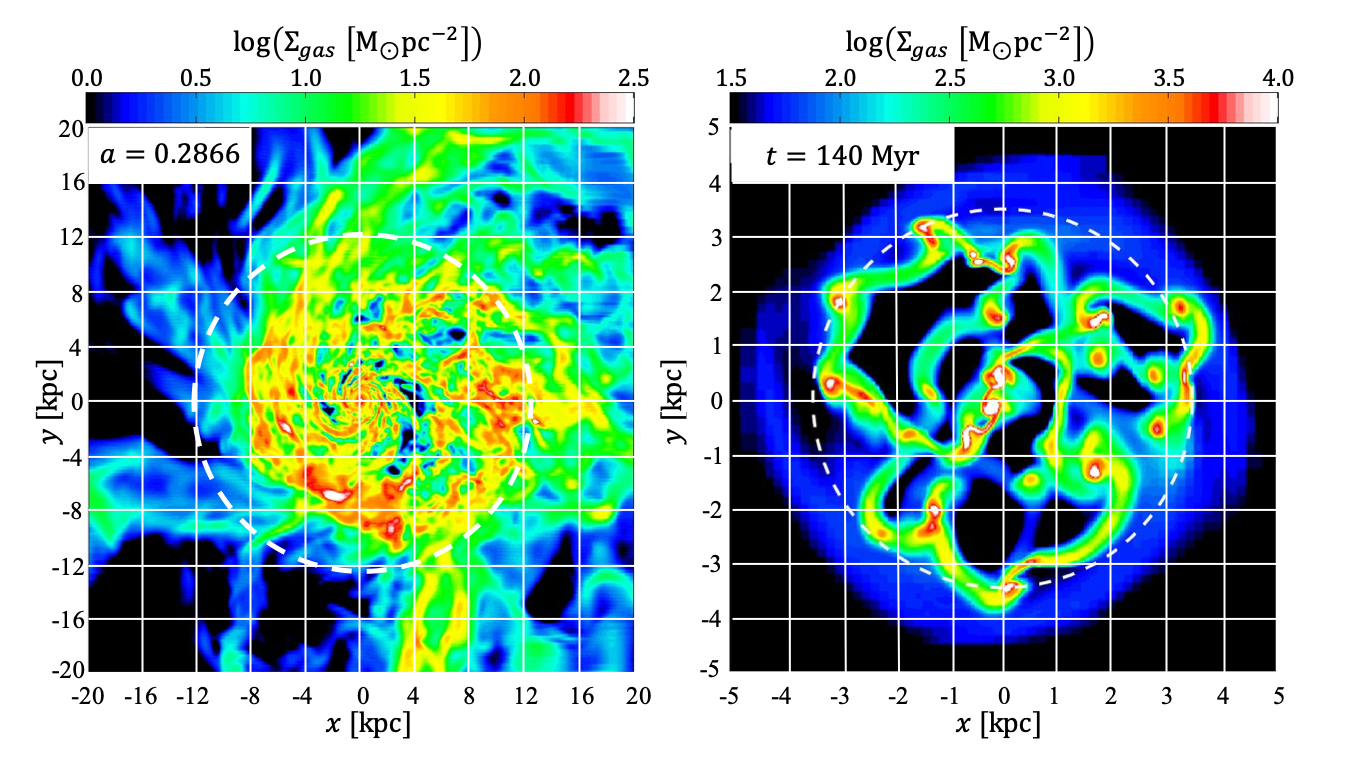}
\end{center}
\vspace{-10.0pt}
\caption{Face on images of gas surface density in our cosmological simulation at $z\sim 2.5$ (left), and our isolated galaxy simulation (right). Both images are shortly after the onset of VDI. Dashed circles mark the disc radius, $R_{\rm d}\sim 12\kpc$ and $3.5\kpc$ in the cosmological and isolated simulations, respectively. The underlying white grid is meant to guide the eye. The cosmological simulation image is integrated over $\pm 20\kpc$, and shows that the disc is fed by two streams from the top and bottom, with most clump formation occurring in a thick ring near the radius where the streams impact the disc. The isolated galaxy image is integrated over $\pm 5\kpc$ and shows dense rings which are fragmenting into giant clumps, as expected from Toomre instability.}
\label{fig:face_on}
\end{figure*}


\smallskip
The right-hand panel of \fig{face_on} shows a face-on projection of the gas surface density at $t\sim 140\Myr$ after the start of the simulation, shortly after the disc begins to fragment into giant clumps. The gas mass fraction at this time is $\sim 56\%$. The disc radius is $R_{\rm d}\sim 3.5\kpc$, while clumps appear to form from the fragmentation of dense axisymmetric rings at $R\sim (2-4)\kpc$.

\subsection{Turbulence Decomposition}
\label{sec:turb}

In general, any velocity field can be decomposed into a compressive part, which has the same divergence as the initial field but no curl, and a solenoidal part, which has the same curl as the initial field but no divergence. For a turbulent  (random) three-dimensional velocity field in equipartition the solenoidal mode has twice as much power as the compressive mode \citep{Federrath.etal.10}, because the former consists of two dimensional vortices while the latter consists of one dimensional shocks and expansion waves. In general, the ratio of energies in each mode can deviate from this value depending on the driver of the turbulence and the local dynamics. At a given turbulent Mach number, the width of the density distribution increases with increasing compressive to solenoidal ratio \citep{Federrath.etal.08}, which contributes to enhanced SFR in regions of excessive compressive modes \citep[e.g.][]{Federrath.etal.12,Renaud.etal.15}. Of particular relevance to Toomre instability, if the velocity dispersion appearing in 
$Q\propto \sigma \kappa/\Sigma$ consists of primarily compressive modes, in particular converging 
flows, then a high value of $\sigma$ would not imply gravitational stability but may actually trigger gravitational collapse. We therefore wish to examine whether 
clump formation is associated with regions of excessive compressive modes. 

\smallskip
There are several methods to numerically evaluate the ratio of compressive to solenoidal modes in a given region. Most studies of the turbulent ISM employ a decomposition in Fourier space. However, such a decomposition can only be used to estimate the total power in each mode in a given region and cannot easily be used to study the spatial correlations of the mode ratios, because the resulting compressive and solenoidal velocity fields are not locally orthogonal in real-space \citep{Brunt.Federrath.14}. To get around this, one can divide the full volume into sub-regions of a given size, perform a Fourier analysis within each sub-region by erroneously assuming it to be periodic rather than part of the full volume, and evaluate the local mode ratios \citep{Vazza.etal.17}. However, here we opt for a fully local approach in real-space. A rigorous local approach is to decompose the shear tensor, the tensor of first derivatives of the velocity field, around every point into three components representing compression/expansion, solid body rotation, and shear flow. We will discuss this approach and its relation to the Fourier decomposition in an upcoming follow-up paper (Ginzburg et al., in preparation). In this work, we instead take a simplified approach employed in previous studies of the compressive to solenoidal ratio in galaxy-scale simulations \citep{Hopkins.etal.13,Renaud.etal.14,Renaud.etal.15,Renaud.etal.15b,Grisdale.etal.17}. We estimate the power in compressive modes at a given point by $\sigma_{\rm c}^2 \equiv |{\bf{\nabla}}\cdot {\bf{v}}|^2 \Delta^2$, the power in solenoidal modes by $\sigma_{\rm s}^2\equiv |{\bf{\nabla}}\times {\bf{v}}|^2 \Delta^2$, and the total turbulent power as $\sigma_{\rm tot}^2\equiv \sigma_{\rm c}^2 + \sigma_{\rm s}^2$. Here, $\Delta$ is the length scale over which the divergence and curl of the field are computed, though this cancels out when taking the ratio. While $\sigma_{\rm c}$ and $\sigma_{\rm s}$ as defined above are not strictly speaking `turbulence', their ratio has been found to be a good proxy for the ratio of turbulent modes. Following \citet{Renaud.etal.14}, we evaluate these on a uniform grid with a grid spacing chosen to be at least 8 times larger than the native simulation resolution, because the curl can artificially dissipate on smaller scales \citep{Federrath.etal.11}. For both cosmological and isolated simulations, we deposit the gas onto a uniform grid with cell size $\Delta=200\pc$ using a cloud-in-cell interpolation, and evaluate derivatives 
using a three-point centered finite difference stencil. Using $\Delta=400\pc$ or a two-point finite difference stencil does not qualitatively change our results. 

\subsection{Clump Identification}
\label{sec:clumps}

We are primarily interested in correlating the ratio of compressive to solenoidal power with the locations of protoclumps undergoing gravitational collapse. We identify protoclumps following the same method employed in \citet{Inoue.etal.16} in their study of Toomre $Q$. We begin by identifying clumps in the simulations using the method detailed in \citet{Mandelker.etal.17}. We deposit the 3D density in a uniform grid with cell size $\Delta_{\rm cl}$, and then smooth it with a spherical Gaussian with full-width-at-half-maximum $W$. We then define $\delta\equiv (\rho-\rho_{_{\rm W}})/\rho$, with $\rho$ and $\rho_{_{\rm W}}$ the unsmoothed and smoothed density fields, respectively. We do this separately for the density of stars and of cold gas ($T<1.5\times 10^4\K$) plus young stars ($<100\Myr$), and take the larger of the two $\delta$ values in each cell. We define clumps as connected regions containing at least $8$ cells with $\delta>\delta_{\rm min}$. For the cosmological simulation, we adopt $\Delta_{\rm cl}=70\pc$ ($2-4$ times the maximal AMR resolution), $W=2.5\kpc$, and $\delta_{\rm min}=10$, as in \citet{Mandelker.etal.17} and \citet{Inoue.etal.16}. For the isolated simulation, we use $\Delta_{\rm cl}=40\pc$ ($2$ times the maximal AMR resolution), $W=0.5\kpc$, and $\delta_{\rm min}=15$. The different values of $W$ and $\delta_{\rm min}$ used in the isolated simulation were calibrated by eye based on face on images of the clumpy discs, and are due to the more compact and gas rich nature of the isolated discs. 

\smallskip
Clumps that contain at least 10 stellar particles are traced through time based on their stellar particles. For each such clump at a given snapshot, we search for all progenitor clumps in the preceding snapshot, defined as clumps that contribute at least 25\% of their stellar particles to the current clump. If a given clump has more than one progenitor, we consider the most massive one as the main progenitor. If a clump in snapshot $i$ has no progenitors in snapshot $i-1$, we search the previous snapshots back to two disc crossing times before snapshot $i$. If no progenitor is found in this period, snapshot $i$ is declared the initial, formation time of the clump. We then trace all clumps back in time one additional timestep to find the location of the protoclump prior to collapse. We do so based on the center-of-mass velocity of the clump at its initial time, in cylindrical coordinates in the disc frame. Since we have $\sim 30$ snapshots per disc-orbital time in both simulations, this extrapolation of the velocity field is reasonable. In what follows, we limit our analysis to clumps with initial mass $M_{\rm c}>10^7\msun$, maximal mass over their lifetime $M_{\rm c,max}>10^8\msun$, formation site within $\pm 1.5\kpc$ of the disc midplane, and survival time of at least one free-fall time, $t_{\rm ff}\equiv \pi/2[R_{\rm c}^3/(GM_{\rm c})]^{1/2}$, with $M_{\rm c}$ the clump baryonic mass and $R_{\rm c}$ the radius of a sphere with the same volume as the clump.

\section{Results}\label{sec:results}

\begin{figure}
\center
\includegraphics[width=0.47 \textwidth]
{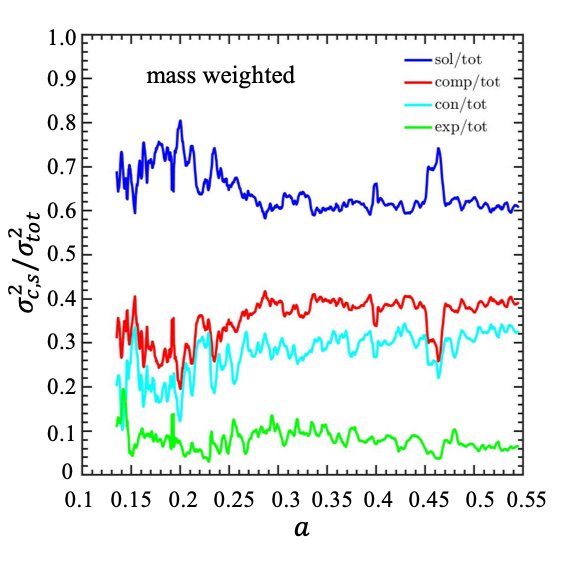}
\caption{Fraction of total turbulent power in various modes in our cosmological simulation. We show as a function of the cosmic expansion factor, the mass-weighted average turbulent power within a cylinder of radius $R_{\rm d}$ and half-thickness $1.5\kpc$. The blue (red) line is the fraction of total power in solenoidal (compressive) modes. The cyan and green lines further divide the compressive power into converging and expanding flows, respectively. See text for details. On average, $\sim 60\%$ of the energy is in solenoidal modes, close to the values of $\sim 2/3$ expected for equipartition. Among the $\sim 40\%$ of the total energy in compressive modes, most of this is in converging rather than expanding flows.
} 
\label{fig:global}
\vspace{-3mm}
\end{figure} 
\begin{figure}
\begin{center}
\includegraphics[trim={0.2cm 0.0cm 0.1cm 0.1cm}, clip, width =0.49 \textwidth]{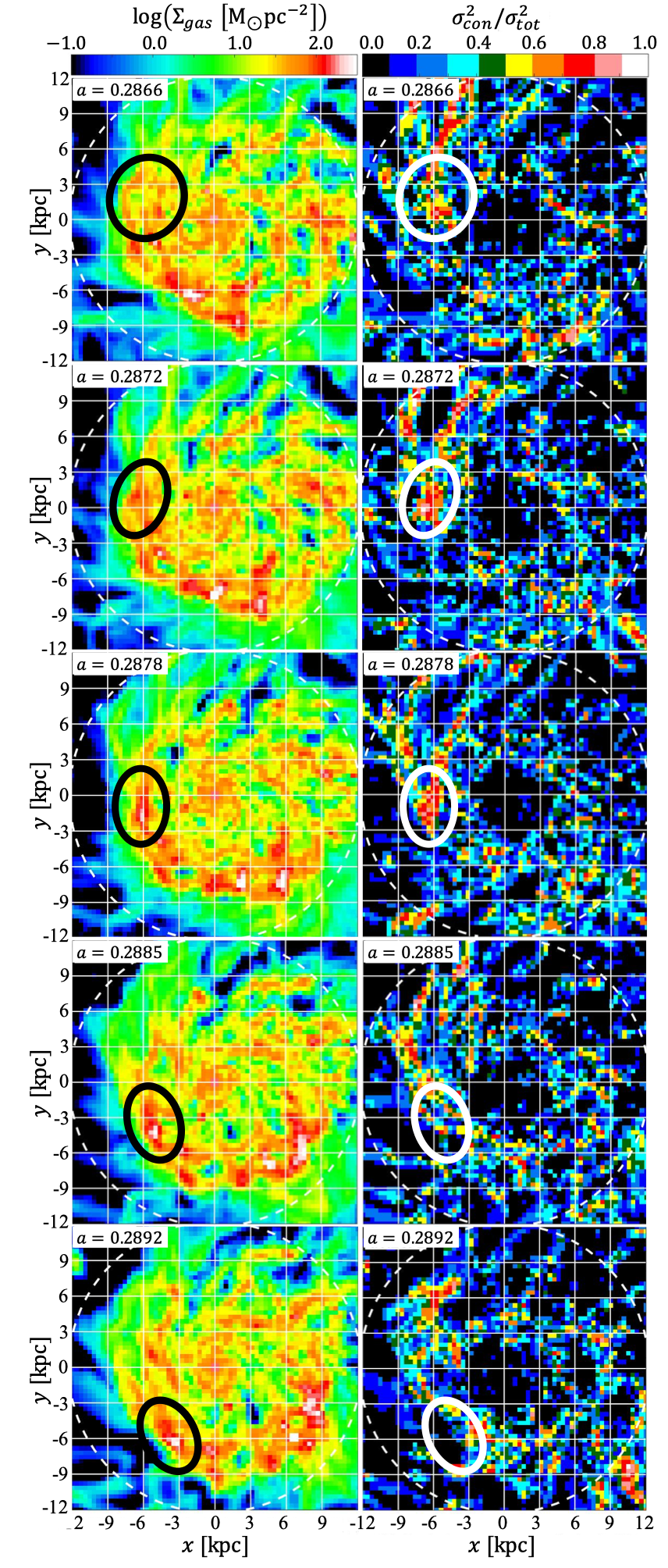}
\end{center}
\vspace{-10.0pt}
\caption{Clump formation in the cosmological simulation. We show maps of gas surface density (left) and the fraction of total turbulent energy in converging motions (compressive modes with negative divergence, right). The maps have pixel sizes of $200\pc$ and are integrated over $\pm 1.5\kpc$. We show the evolution from $z\sim 2.49$ on the top to $z\sim 2.46$ on the bottom, with roughly $\sim 9\Myr$ in between snapshots. We mark a forming clump from its protoclump phase with a balck (white) circle in the left (right) columns. The turbulence in the protoclump region is dominated by converging motions.
}
\label{fig:protoclump}
\end{figure}

\Fig{global} shows the fraction of total turbulent energy in various modes (see \se{turb}) in our cosmological simulation, as a function of cosmic expansion factor, $a$. For each mode, $\sigma^2$ represents the mass-weighted average 
over 
a cylinder of radius $R_{\rm d}$ and half-thickness $1.5\kpc$, 
where most clump formation occurs and where the gas densities are high. At $z<3$ ($a>0.25$), $\sim 60\%$ of the turbulent energy is in solenoidal modes (blue curve) while the remaining $\sim 40\%$ is in compressive modes (red curve). 
This represents a slight ($\sim 10\%$) excess of compressive modes compared to equipartition, where the expected fractions are $\sim 2/3$ and $\sim 1/3$.
At earlier times, there is a large excess of solenoidal modes, presumably associated with the growth of angular momentum in the galaxy prior to the formation of the disc. We further show the energy fraction in converging (cyan) and expanding motions (green), defined as compressive modes with negative and positive divergence, respectively. We find that $\sim 80\%$ of the energy in compressive modes is associated with converging motions, compared to $\sim 20\%$ in expanding motions. This is consistent with the compressive modes being dominated by regions of 
collapse rather than outflows. 

\smallskip
In the isolated galaxy simulation (not shown), we find $\sim 70\%$ of the energy in solenoidal modes and $\sim 30\%$ in compressive modes at all times, very close to equipartition. Furthermore, most of the energy in compressive modes is actually associated with expanding rather than converging motions. 
The cosmological simulation thus exhibits an excess in compressive modes associated with collapse compared to the isolated galaxies. This excess is small when averaged over the entire disc, and the compressive modes never dominate the global turbulence, as they may do following major-mergers triggering galaxy-wide starbursts \citep{Renaud.etal.14}. This is further supported by the fact that the volume-weighted average values of $\sigma^2$ yield $\sim 68\%$ ($\sim 32\%$) in solenoidal (compressive) modes, 
consistent with equipartition, with most of the compressive modes being associated with expanding motions. 

\smallskip
In \fig{protoclump}, we examine the correlation between compressive modes, specifically converging motions, and clump formation. We show a time sequence of our cosmological simulation spanning $\sim 36\Myr$ from $z\sim (2.49-2.46)$. We show the surface mass density of gas on the left, and the ratio of the (mass-weighted) energy in converging motions to total turbulent energy along the line-of-sight on the right. Each image has been integrated over $\pm 1.5\kpc$, as in \fig{global}, and has a pixel size of $200\pc$ (see \se{turb}). In each panel, we highlight the position of a forming clump. This clump is first identified at expansion factor $a=0.2872$ (second row), and the position marked at $a=0.2866$ (top row) is the extrapolated protoclump position (see \se{clumps}). In the protoclump patch, converging motions comprise $\sim 70\%$ of the total turbulent energy, compared to $\sim 17\%$ expected in equipartition where the energy in compressive modes is evenly split between converging and expanding flows. This large excess of energy in converging flows is maintained for the first $\sim 20\Myr$ of the clump lifetime, after which it reduces to the equipartition value. It is worth noting that after collapse, the clumps become dominated by rotation \citep{Ceverino.etal.12} contributing to additional solenoidal modes inside the clump. This same evolutionary sequence of clump formation was shown in \citet{Inoue.etal.16}, Fig. 10, where it was shown that $Q\sim 5$ in the protoclump region, and only reached $Q\lsim 1$ after $\sim 20\Myr$. We thus see that excessive compressive modes, and in particular converging flows, can trigger clump formation in regions where $Q\gg 1$. 

\begin{figure}
\center
\includegraphics[trim={0.2cm 0.0cm 0.2cm 0.0cm}, clip, width=0.49 \textwidth]
{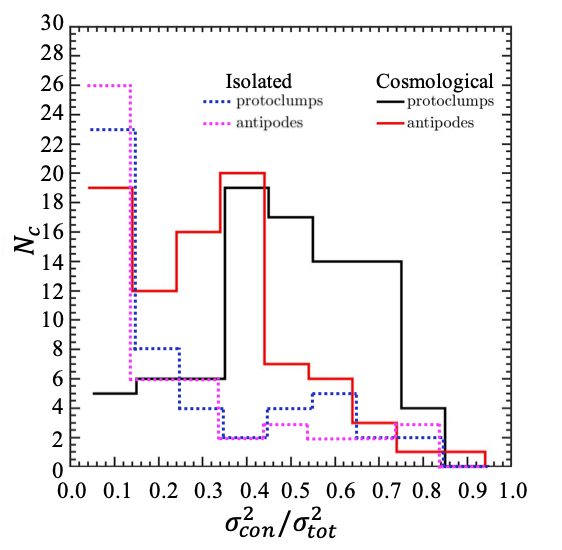}
\caption{
Distribution of the fraction of total turbulent energy in converging motions in protoclumps (solid black and dotted blue) and their antipodal regions (solid red and dotted magenta, see text) in the cosmological (solid lines) and isolated galaxy simulations (dotted lines). The energies are computed in a circular aperture of radius $500 \pc$ centered on the protoclumps or their antipodes, extending $\pm 1.5\kpc$ about the disc midplane. In the isolated simulation, the protoclumps and the antipodes have very similar distributions, sharply peaked at a ratio of $\lsim 0.1$, slightly below the value expected for equipartition. In the cosmological simulation, the protoclumps have typical values of $\sim 0.6-0.7$ while the antipodes have typical values of $\sim 0.3$. Excessive energy in converging flows is thus associated with clump formation in the cosmological simulation, but not in the isolated galaxy simulation. 
} 
\label{fig:histogram}
\vspace{-3mm}
\end{figure} 

\smallskip
One may worry that it is not the excessive converging motions that trigger clump formation, but rather that the already collapsing protoclump induces excessive converging flows. We address this, and systematically compare clump formation in the cosmological and isolated galaxy simulations, in \fig{histogram}. We show the distribution of the ratio of energy in converging motions to the total turbulent energy in protoclump patches 
in the cosmological and isolated galaxy simulation as solid black and dotted blue histograms, respectively. 
The energies are mass-weighted averages within a circular aperture of radius $500\pc$ around the center of the protoclump, and within $\pm 1.5\kpc$ of the disc midplane as in \figs{global} and \figss{protoclump}. For each protoclump, we also compute the energy ratio in the same aperture but centered on the antipodal region, namely the region at the same galactocentric radius but on the opposite side of the disc, $180^\circ$ from the protoclump. The distribution of energy ratios in these antipodal regions are shown in 
solid red and dotted magenta for the cosmological and isolated simulations, respectively.

\smallskip
In the isolated simulation, the ratio of energy in converging flows to total turbulent energy is strongly peaked at $\lsim 0.1$ in both the protoclumps and their antipodes, slightly below the value expected for equipartition. This is consistent with the global disc average where most of the compressive energy is in expanding flows, presumably due to feedback.
On the other hand, in the cosmological simulations, the protoclump regions are strongly biased towards large values of converging flows, typically $\sim 0.6-0.7$ of the total turbulent energy. The converging-to-total energy ratios in the antipodes have typical values of $\sim 0.3$, much lower than in the protoclumps though still high compared to the isolated simulation and the equipartition value. 

\section{Discussion and conclusions} \label{sec:concl}

We have shown that violently unstable discs at $z\sim 2$ in cosmological simulations exhibit an excess in compressive turbulence compared to the equipartition value (\fig{global}). This excess is not seen in isolated galaxy simulations. 
Furthermore, our results show that there is a qualitative difference between the cosmological and isolated simulations in terms of the correlation between clump formation and excessive compressive modes, specifically converging motions with negative divergence. In cosmological simulations, converging flows account for up to $\sim 70\%$ of the total turbulent energy in protoclump regions, which can trigger clump formation even in regions where $Q\gg 1$ (\fig{protoclump}, compare to Fig. 10 in \citealp{Inoue.etal.16}). In isolated simulations, on the other hand, protoclump regions appear to be in equipartition, with converging motions accounting for $\sim 1/6$ of the total turbulent energy (\fig{histogram}). This suggests that the excessive energy in converging flows seen within protoclumps in the cosmological simulation is of an external origin; it is unlikely caused by the collapse of the clump itself, or else we would have seen a similar signature in the isolated galaxy simulation.

\smallskip
We conclude that the formation of giant clumps during VDI can be triggered by converging flows associated with excessive compressive turbulence. This is a non-linear process and is thus not directly related to the local value of Toomre $Q$. The question is what causes such an excess. Since it is not seen in isolated galaxy simulations, it must be external in origin. One possibility is that it is induced by intense accretion from cold streams flowing along the cosmic web. This is known to be an important aspect of VDI \citep{Dekel.etal.09b}, and has been shown to be capable of inducing turbulence in discs at levels comparable to both supernova feedback and gravitational instabilities \citep{Ginzburg.etal.22}, depending on the density distribution in the incoming streams. As can be seen by comparing \figs{face_on} and \figss{protoclump}, the region with a large excess of energy in converging motions where clumps tend to form correlates with the region where inflowing streams join the disc. Another possibility is that the excessive compressive modes are induced by an excess in fully compressive tidal forces. This was shown to be the case in major-merger simulations, where the tidal forces induced by the merger were fully compressive across wide regions, leading to a large excess in compressive turbulence which then triggered a starburst \citep{Renaud.etal.14,Renaud.etal.15}. Similar tidally-induced clumps were found to form during major mergers in cosmological simulations at $z\sim 8$ \citep{Nakazato.etal.24}. The cosmological simulation studied here does not undergo any major mergers, and indeed the global ratio of compressive to solenoidal motions is only modestly greater than equipartition. However, the chaotic nature of the galaxy environment, including accretion of dense gas, minor mergers, pre-existing clumps and other perturbations, may induce local tidal forces which are fully compressive, leading to local excesses in converging flows and clump formation. Finally, we cannot at this time rule out a third possibility, that the difference is caused by the different feedback schemes implemented in the cosmological and idealized simulations, rather than an external trigger. 

\smallskip
Distinguishing these possibilities, and clarifying the origin of excessive compressive modes in 
VDI galaxies more broadly, will be the focus of an upcoming paper (Ginzburg et al., in preparation), where we examine a larger sample of comsological and isolated galaxy simulations in more detail.

\section*{Acknowledgements}
We thank Frank van den Bosch, Christoph Federrath, Shigeki Inoue, and Florent Renaud for very helpful and educational discussions. NM acknowledges support from Israel Science Foundation (ISF) grant 3061/21. OG is supported by a Milner fellowship and, along with AD, by ISF grant 861/20. DC is a Ramon-Cajal Researcher and is supported by the Ministerio de Ciencia, Innovacion y Universidades (MICIU/FEDER) under research grant PID2021-122603NB-C21. MRK acknowledges support from the Australian Research Council through its Laureate Fellowship program, award FL220100020, and from access to computational resources supported by the Australian Government's National Collaborative
Research Infrastructure Strategy and distributed through the National Computational Merit Allocation Scheme, award jh2.

\bibliographystyle{mnras}
\bibliography{references}

\bsp

\label{lastpage}

\end{document}